\documentclass[prb,twocolumn,showpacs,showkeys,floatfix,superscriptaddress]{revtex4}

\usepackage{graphicx}% Include figure files
\usepackage{dcolumn}% Align table columns on decimal point
\usepackage{bm}% bold math
\usepackage[T1]{fontenc}

\begin{document}

\preprint{APS/123-QED}

\title{Room temperature magnetism in LaVO$_3$/SrVO$_3$ superlattices by geometrically confined doping}% Force line breaks with \\

\author{U. L\"uders}
\affiliation{%
CRISMAT, UMR CNRS-ENSICAEN 6508,
6 boulevard Maréchal Juin, 14050 Caen cedex 4, France
}%
\author{W. C. Sheets}%
\affiliation{%
CRISMAT, UMR CNRS-ENSICAEN 6508,
6 boulevard Maréchal Juin, 14050 Caen cedex 4, France
}%
\author{A. David}%
\affiliation{%
CRISMAT, UMR CNRS-ENSICAEN 6508,
6 boulevard Maréchal Juin, 14050 Caen cedex 4, France
}%
\author{W. Prellier}
\affiliation{%
CRISMAT, UMR CNRS-ENSICAEN 6508,
6 boulevard Maréchal Juin, 14050 Caen cedex 4, France
}%
\author{R. Fr\'esard}%

\affiliation{%
CRISMAT, UMR CNRS-ENSICAEN 6508,
6 boulevard Maréchal Juin, 14050 Caen cedex 4, France
}%

\date{\today}

\begin{abstract}

Based on the Hubbard model of strongly correlated systems, a reduction in the bandwidth of the electrons can yield a substantial change in the properties of the material. One method to modify the bandwidth is geometrically confined doping, i.e. the introduction of a (thin) dopant layer in a material. In this paper, the magnetic properties of LaVO$_3$/SrVO$_3$ superlattices, in which the geometrically confined doping is produced by a one monolayer thick SrVO$_3$ film, are presented. In contrast to the solid solution La$_{1-x}$Sr$_x$VO$_3$, such superlattices have a finite magnetization up to room temperature. Furthermore, the total magnetization of the superlattice depends on the thickness of the LaVO$_3$ layer, indicating an indirect coupling of the magnetization that emerges at adjacent dopant layers. Our results show that geometrically confined doping, like it can be achieved in superlattices, reveals a way to induce otherwise unaccessible phases, possibly even with a large temperature scale.

\end{abstract}

\maketitle

The physics of doped Mott insulators embraces a wealth of fascinating properties such as high temperature superconductivity \cite{Bed86,Rav92}, colossal magnetoresistance \cite{Hel93,Mai95} and stripe formation \cite{Sac95}, among others. While these properties are linked with mixed valent perovskite-like systems containing Cu, Mn or Ni, systems involving V ions have not been associated previously with such spectacular effects. Indeed, the phase diagram of the solid solution La$_{1-x}$Sr$_x$VO$_3$ is quite simple compared to the above cited materials. LaVO$_3$ is a Mott insulator, showing an antiferromagnetic transition at 143K and a structural transition at 141K \cite{Miy03}. Upon doping with Sr, this antiferromagnetic transition persists up to a Sr concentration x $\approx$ 0.2 \cite{Say75,Ina95,Miy00}, where an insulator to metal transition occurs. For higher x, La$_{1-x}$Sr$_x$VO$_3$ becomes a non-magnetic metal, yet showing a non-Fermi liquid behavior of the resistivity. Its T$^{3/2}$ dependence indicates strong magnetic fluctuations \cite{Ina95}, suggesting that the system is on the verge of a magnetic transition, which still remains to be evidenced. 

In fact, such a magnetic transition is expected on the basis of the extensive theoretical work devoted to strongly correlated systems, especially on the basis of multiband Hubbard-type models (see \cite{Ima98,Vol97,Dag01} for reviews) or by means of electronic structure calculations with dynamical mean-field theory \cite{Kot06}. Among the examined instabilities of the weakly interacting paramagnetic state, a ferromagnetic ground state has been the focus of numerous studies, following the pioneering paper by Lacroix-Lyon-Caen and Cyrot \cite{Lac77}. Since then, it has been clarified that the underlying non-interacting model plays a key role when assessing the instabilities, be it a doubly degenerate Hubbard Model, or a model involving two degenerate $e_g$ or $t_{2g}$ orbitals. For instance, evidence for a ferromagnetic ground state in the vicinity of three quarter filling (the expected relevant density of charge carriers in our metallic layers) has been put forward by means of exact diagonalization for the doubly degenerate Hubbard Model for arbitrary interactions strengths $U$ and $J_H$ by Romano et al. \cite{Rom08}, or above a critical ratio of the interaction strength $U$ to the band width $W$ for the two-dimensional Hubbard Model involving two degenerate $t_{2g}$ orbitals by means of slave boson saddle point approximation \cite{Fre02}. As the latter model should be relevant to La$_{1-x}$Sr$_x$VO$_3$, the experimental observation of a metallic
paramagnetic phase indicates a value of U/W smaller than the critical one for the solid solution. Yet, if one experimentally succeeded in increasing U/W, for example by reducing W by means of geometrically confined doping (GCD), a ferromagnetic phase could emerge.

\begin{figure}
\includegraphics[width = 0.5\textwidth]{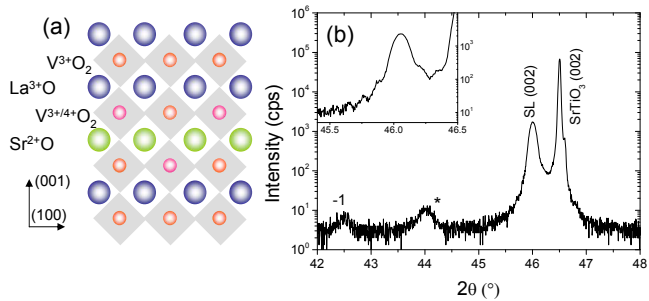}% Here is how to import EPS art
\caption{\label{Figure1} (color online) (a) Sketch of the superlattice in the ac plane. (b) $\theta$-2$\theta$ scan of a high pressure m = 6 sample. A superlattice satellite peak is indicated by its order number. The reflection indicated by a star is from the sample holder. Inset: Zoom on the superlattice Bragg peak of a m = 5 sample.}
\end{figure}

Our approach to reduce the bandwidth of the dopant electrons in La$_{1-x}$Sr$_x$VO$_3$ is introducing doping sublayers of SrO in (001) oriented layers of LaVO$_3$ by growing superlattices of LaVO$_3$/SrVO$_3$. The SrVO$_3$ layers have a nominal thickness of one unit cell, whereas the LaVO$_3$ has a thickness of 6 or less unit cells. Such samples can be considered as LaVO$_3$, in which certain LaO subplanes are replaced by a SrO subplane, leading to a doping of the adjacent VO$_2$ subplanes (see Figure \ref{Figure1}(a)) due to the difference in oxidation state of La and Sr. The insulating character of the LaVO$_3$ layers confines the doped charge carriers in the vicinity of the SrO subplane, imposing a two-dimensional (2D) character to the 3d t$_{2g}$ bands, which reduces their bandwidth.

To our knowledge, no magnetic measurements were published on such GCD systems of t$_{2g}$ electrons \endnote{Extensive work was done on GCD systems of e$_g$ electrons, as for example LaMnO$_3$/SrMnO$_3$ superlattices. However, in these systems the bandwidth of the e$_g$ electrons is smaller and the magnetic character has been observed already in the solid solution.}. While superlattices of LaTiO$_3$/SrTiO$_3$ have been successfully prepared \cite{Oht02,Shi04} and the artificially imposed charge modulation of the Ti ions was demonstrated, no magnetic data were published. Magnetism was shown recently at polar discontinuous interfaces between non-magnetic materials \cite{Bri07} at much lower temperature. Single (001) interfaces of LaVO$_3$/SrVO$_3$ were realized experimentally \cite{hot07} and considered theoretically \cite{Jac08}, agreeing that this interface is insulating, but the theoretically predicted magnetic character was not verified. Even though the superlattices in our study contain the same interfaces, two reasons suggest not to discuss our results in this framework: first, the interface was shown to be insulating, while GCD samples show conducting behavior, thus the observed effects should be understood based on doping. Second, magnetism was also observed in (110) GCD samples, in which the interfaces do not exhibit a polar discontinuity \cite{hot07}.  

The samples were prepared by Pulsed Laser Deposition on SrTiO$_3$ (001) substrates, details are described elsewhere \cite{She07}. Single films of both insulating LaVO$_3$ and metallic SrVO$_3$ show magnetic and transport properties similar to their respective bulk counterparts. The superlattices consist of repetitions of a (LaVO$_3$)[m]/(SrVO$_3$)[1 unit cell] bilayer, where m is the thickness of the LaVO$_3$ layer and was varied between m = 2 to 6 unit cells. The deposited number of layers was verified by the analysis of the superlattice satellite peaks in the X-ray diffraction (XRD) pattern, the separation in angle of which is directly related to the thickness of the bilayers. The number of repetitions was adjusted so that the total thickness of all samples is around 110nm. A single film of the solid solution with x = 0.15 was made by alternating the deposition of the two materials with submonolayer period in this stochiometric ratio. This sample will serve as a reference sample.
 
The structural properties were measured by XRD using a Seifert 3000P diffractometer with the Cu K$_{\alpha1}$ wavelength of 1.5406\AA. Magnetic measurements were done in a Magnetic Property Measurement Device SQUID (Quantum Design) with the magnetic field applied in the plane of the film along a (100) direction. No magnetic in-plane anisotropy was observed. As the magnetization of the superlattices is small, misinterpretations due to magnetic artifacts must be avoided. For this reason, the magnetization of bare SrTiO$_3$ substrates was measured at different temperatures. On top of the expected diamagnetic behavior, a small hysteresis with a magnetization at high field on the order of 10$^{-5}$ emu was measured, which is most likely caused by magnetic impurities. This magnetization, if erroneously attributed to the thin film and therefore normalized on a typical film volume, would correspond to a magnetization of around 15emu/ccm. Therefore, a hysteresis loop of a sample with a saturation magnetization of 15emu/ccm or less cannot be attributed unequivocally to the thin film, and is probably due to artifacts within the substrate. To account for the diamagnetic contribution of the substrate to the total magnetization, the high field magnetization slope of the hysteresis loops was fitted and subsequently subtracted from the data.

\begin{figure}
\includegraphics[width = 0.4\textwidth]{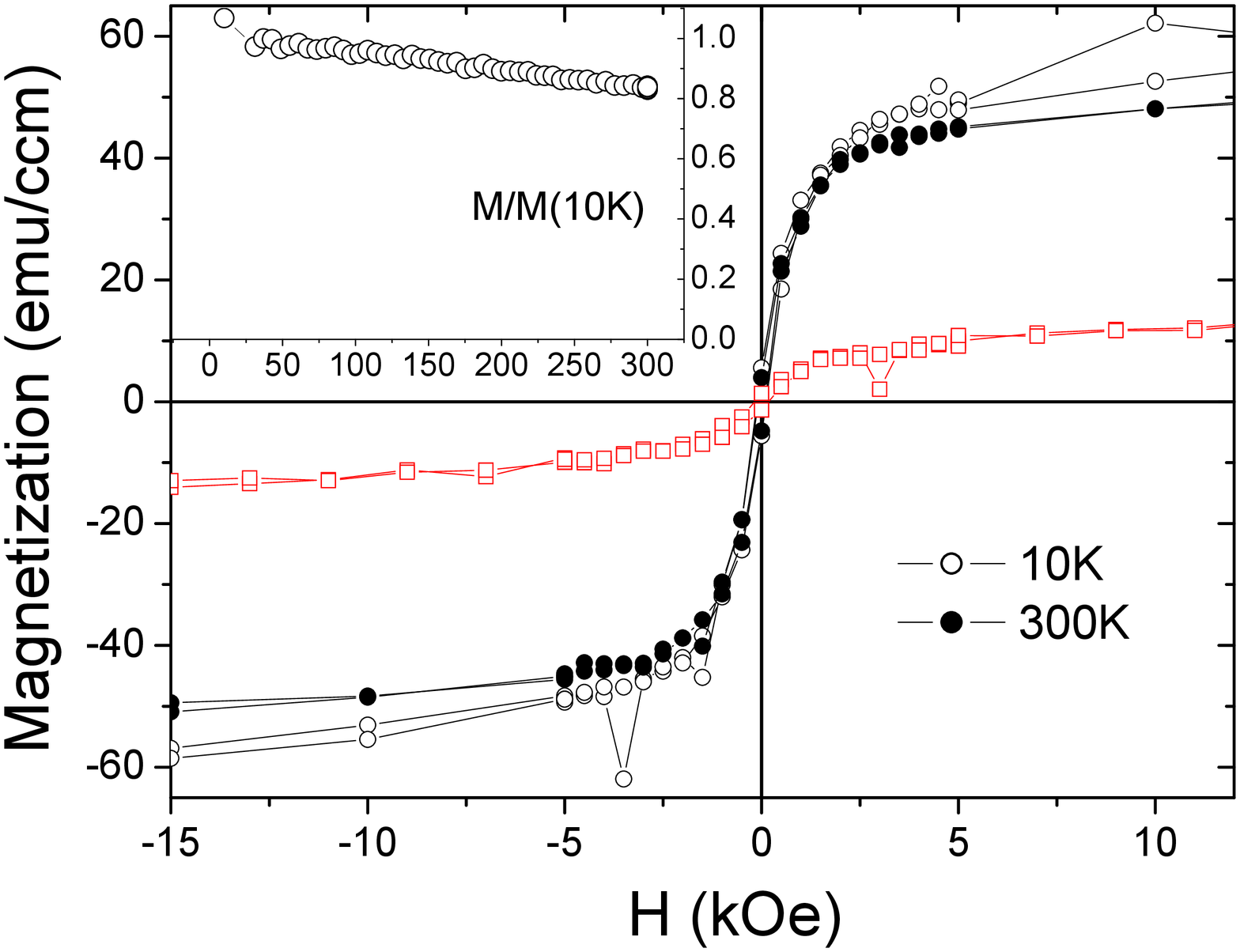}% Here is how to import EPS art
\caption{\label{Figure2} (color online) Magnetization $vs$ magnetic field for a high pressure m = 6 superlattice (circles) at 10K and 300K, and the solid solution single layer (squares) at 10K. Inset: Normalized magnetization of the same suo vs temperature in an applied field of 500Oe.}
\end{figure}

To investigate the influence of the localization of the electrons on the properties of the superlattices, two different series were prepared in vacuum at different pressures. The first series prepared in a pressure of 4 x 10$^{-5}$mbar will be referred to hereafter as the high pressure series. The low pressure series was grown in a pressure of 1 x 10$^{-5}$mbar. In-plane transport measurements of the two different series show a metallic character at room temperature for any value of m. Yet, the value of the resistivity varies between the two series: the high pressure series shows a room temperature resistivity of around 1 to 10m$\Omega$cm, while for the low pressure series a resistivity of around 0.1 to 3m$\Omega$cm was observed. Note that transport properties on similar superlattices were published recently \cite{She09} showing insulating behavior for the m > 4 superlattices. These superlattices were deposited at even higher pressure, illustrating the criticality of this parameter for the properties of the samples. First evidence of the layered structure of our samples is provided by out-of-plane resistance measurements, which were carried out for the high pressure series. They revealed a temperature dependence typical of insulators, in contrast to the in-plane measurements.

A typical result of out-of-plane XRD measurements at room temperature is shown in Figure \ref{Figure1} for the high pressure m = 6 sample. Except of indications of an increasing quality of the samples with decreasing deposition pressure, spectra for both series do not show significant differences. Combined with in-plane XRD measurements, a cube-on-cube epitaxy of the superlattices on the substrate is deduced. The perovskite structure presents a small tetragonal distortion (a$_c$/c = 0.388nm/0.395nm = 0.98, where a$_c$ is the in-plane lattice parameter of the pseudo-cubic representation and c the out-of-plane lattice parameter). The measured a$_c$ corresponds to a value located between the lattice parameter of bulk LaVO$_3$ (0.392nm) and SrVO$_3$ (0.382nm), and is slightly smaller than the substrate value. 

An important issue is the morphology of the SrO doping layers. The presence of superlattice satellites and Laue fringes at the Bragg peak of the superlattice (002) reflection indicates a layered structure with an accomplished supercell and smooth interfaces, and therefore indirectly supports the continuous character of the dopant layer. Moreover, high-resolution electron microscopy images were taken of samples with 3 unit cell thick SrVO$_3$ layers \cite{She09}, showing continuous SrVO$_3$ layers with an interface roughness not exceeding one monolayer. Such evidence is a strong indication that we have achieved the deposition of continuous 2D SrO subplanes.

\begin{figure}
\includegraphics[width = 0.4\textwidth]{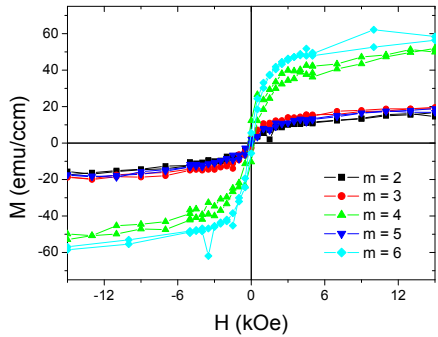}% Here is how to import EPS art
\caption{\label{Figure3} (color online) Magnetization $vs$ magnetic field for the high pressure series with m ranging from 2 to 6 at 10K.}
\end{figure}

In Figure \ref{Figure2} the hysteresis loops of two films with the same nominal composition are compared, but with different dimensionality of the doping: a high pressure m = 6 superlattice (GCD) and the La$_{0.85}$Sr$_{0.15}$VO$_3$ solid solution single layer (3-dimensional doping). For the solid solution film, the composition of which is in the range of the bulk antiferromagnetic insulating phase, a small hysteresis is found at low temperature. As its magnetization is in the range of 15emu/ccm, it can be attributed to substrate impurities as discussed previously. Regarding the superlattice, the hysteresis loop shows a total magnetization that is more than three times higher. Thus, the superlattices are magnetically ordered due to the introduction of the GCD. The comparison with the solid solution film demonstrates that the magnetization in the superlattices can neither be growth induced, nor attributed to impurities or other artifacts due to the treatment of the sample or the data. 

\begin{figure}
\includegraphics[width = 0.4\textwidth]{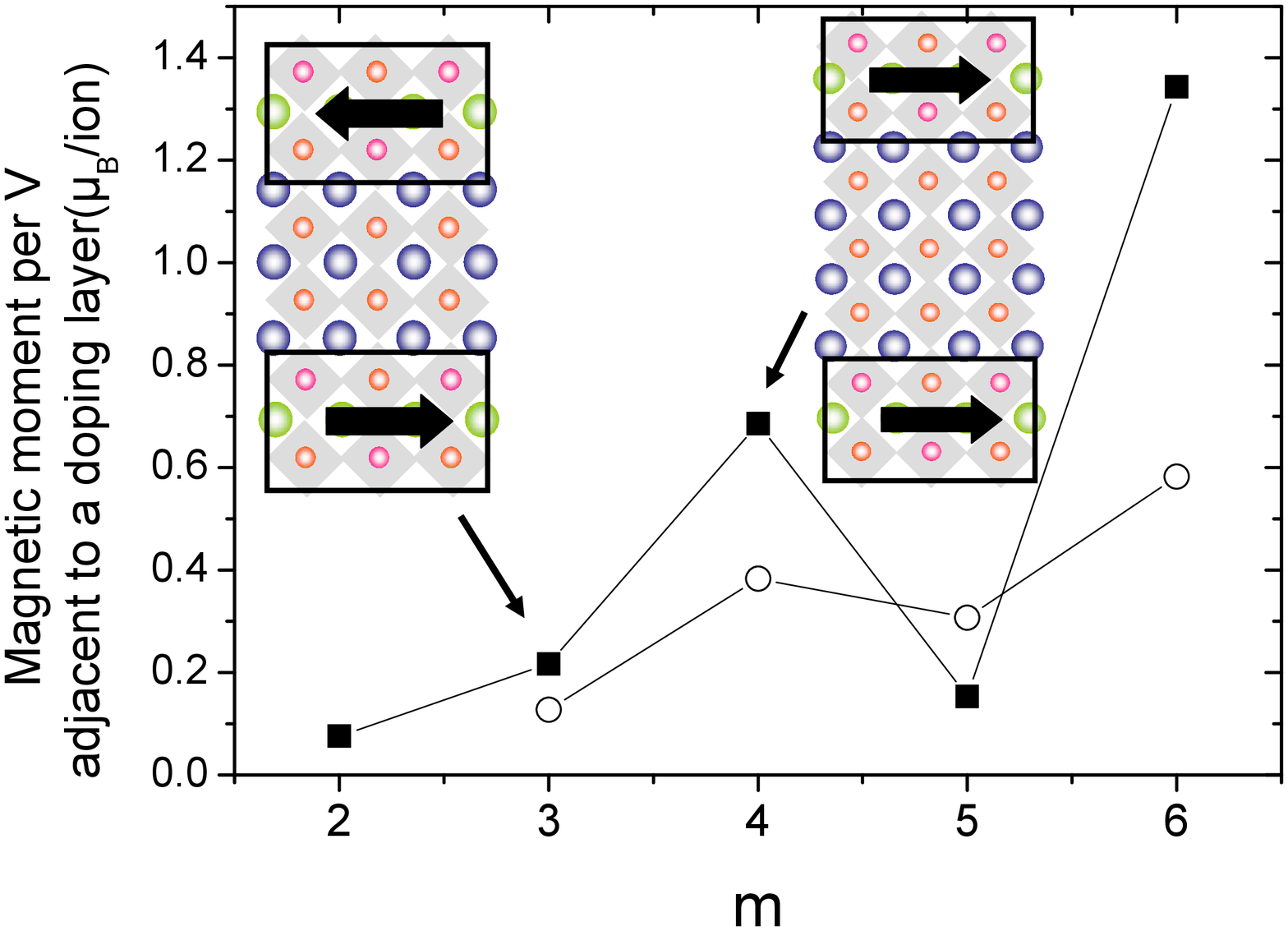}% Here is how to import EPS art
\caption{\label{Figure4} (color online) Saturation magnetization of the superlattices of the high pressure series (solid squares) and the low pressure series (open circles) $vs$ m at 10K. The sketches in the figure show a possible magnetic structure of the superlattice.}
\end{figure}

The saturation magnetization of the superlattices varies with m. As can be observed in Figure \ref{Figure3} for the high pressure series, superlattices with m = 4 and 6 show a saturation magnetization of around 50emu/ccm, whereas the superlattices with m = 2, 3, and 5 have a low magnetization in the range of 15emu/ccm. The magnetic moment of the same samples along with the low pressure series are summarized in Figure \ref{Figure4}. Both superlattice series exhibit the same behavior, although the amplitude of the magnetization variation for the low pressure series is smaller. 

The above evidenced magnetic order of the superlattices can be understood on the basis of the spatial control of the doping layer. As mentioned above, the two dimensional character of the doping planes leads to a reduction of the bandwidth of the bands crossing the Fermi energy and therefore a stronger localization of the 3d electrons, favoring a ferromagnetic order. Such a bandwidth reduction is robust against various forms of moderate disorder, in particular the possible buckling of the SrO subplanes and their adjacent VO$_2$ subplanes. Consequently, the ferromagnetic order can sustain a certain level of disorder. However, ferromagnetism is not obtained in the strong disorder limit, namely when the film consists of a solid solution with the same nominal Sr concentration. Therefore we can conclude, without having shown explicitly the electron confinement or the exact character of the SrO doping planes, that the magnetization is induced by the layered structure of the sample.

The temperature dependence of the magnetization of the superlattices was measured up to room temperature (see inset of Figure \ref{Figure2}). The observed decrease of the magnetization is very shallow and 80\% of magnetization at 10K remains at room temperature. Therefore, the Curie temperature seems to be far above room temperature, indicating a robust magnetic exchange interaction in the samples. The solid solution and the related compounds show no magnetic long-range order or a substantially lower magnetic transition temperatures, which stresses the opportunities provided by the GCD method.

The magnetic moment per V ion in the mixed valence VO$_2$ layers was extracted from the magnetization measurements under the assumption that each SrO layer accounts for two doped VO$_2$ subplanes. Charge distribution calculations performed for LaMnO$_3$/SrMnO$_3$ superlattices \cite{Ada09} showed that the mixed valence is restricted principally to the two monolayers directly at the interface. The extracted values are shown in Figure \ref{Figure4}. The highest magnetic moment is found for the high pressure m = 6 superlattice with a value of 1.4$\mu_B$/V ion. In a simple model, the VO$_2$ subplanes adjacent to the SrO doping layers will have an equal amount of V$^{3+}$ and V$^{4+}$. The corresponding magnetic moments are 2$\mu_B$ and 1$\mu_B$ per V ion, respectively, resulting in an overall magnetic moment of 1.5$\mu_B$ per V ion in the VO$_2$ subplanes adjacent to the SrO doping layer, which is consistent with the observed value.

The maximum magnetization of the low pressure series is smaller than for the high pressure series. This difference emphasizes the role of localization of the t$_{2g}$ electrons in the superlattices. In the solid solution, magnetic order does not develop because the electrons are mostly itinerant \cite{Ngu95}. Therefore, in the more metallic low pressure series, the enhanced order at the doping layer leads to a suppression of the magnetic moment. Let us note that a variation in oxygen vacancies due to the change in deposition pressure would lead to the opposite effect.

The absence of a macroscopic magnetization for the m = 2 sample is most probably due to the small thickness of the LaVO$_3$ layers, in which case the previously discussed reduction of the bandwidth is less effective, especially when taking the likely leakage of the charge carriers into the single remaining non-doped VO$_2$ subplane into account. For the m $\geq$ 2 superlattices, the m-dependence of the magnetization exhibits a clear even-odd effect, which can not be explained by a variation of the magnetic moments of the magnetic VO$_2$ subplanes, as the addition of one monolayer of LaVO$_3$ only changes the spacing of the GCD zones, but not their inherent character. Instead, we propose an indirect coupling of the magnetic VO$_2$ subplanes resulting in a ferromagnetic or antiferromagnetic alignment depending on the thickness of the separating LaVO$_3$ layer.

This is similar to the theoretical prediction by Jackeli and Khaliullin \cite{Jac08} that the coupling of the magnetic LaVO$_3$/SrVO$_3$ interfaces oscillates with the thickness of the LaVO$_3$ layer. Yet, their result was obtained assuming fully nested Fermi surfaces, and therefore an insulating ground state, which is at odds with the metallic character of our superlattices. Besides, it should be noted that no peculiar temperature dependence of the magnetization in the entire temperature range below room temperature was observed, especially in the vicinity of the Néel temperature of bulk LaVO$_3$. Moreover, the oscillating coupling subsists up to room temperature.

In conclusion, a room temperature magnetization was found in two dimensionally doped LaVO$_3$/SrVO$_3$ superlattices. The emerging magnetic behavior is attributed to the lower bandwidth of the conduction electrons and therefore a direct result of the strongly correlated character of the electrons in this system. The relevance of the bandwidth was indirectly shown by the lower magnetic moment per V ion in low resistivity samples. The observed magnetic moment is consistent with a mixed valence of the VO$_2$ layers adjacent to the SrO doping layers. The magnetization of the superlattices varies with the number of LaVO$_3$ unit cells separating the doping layers, and can be attributed to a form of indirect coupling of the mixed valence VO$_2$ layers, indicating rich physics in this system.

The authors acknowledge fruitful discussions with C. Simon and T.W. Noh, as well as the financial support of CNano, Cefipra and STAR.

\end{document}